\documentclass{article}

\usepackage{amsfonts, amsmath, amssymb, amsbsy}
\usepackage{graphicx}
\usepackage{epstopdf}
\begin{document}

\begin{center}
\textbf{\LARGE S5 1803+78 revisited}
\end{center}

\begin{center}
\textbf{R. Nesci$^1$, A. Maselli$^2$, F.  Montagni$^3$, S. Sclavi$^1$}
\end{center}

\begin{center}
{\it
\noindent $^1$University La Sapienza, Roma, Italy\\
$^2$INAF-IASF, Palermo, Italy \\
$^3$Greve in Chianti Observatory, Italy }
\end{center}

\begin{abstract}
We report on our optical monitoring of the BL Lac object S5 1803+78 from 1996 to 2011. The source showed no clear periodicity, but a time scale  of about 1300 days between major flares is possibly present. No systematic trend of the color index with flux variations is evident, at variance with other BL Lacs.  In one flare, however, the source was bluer in the rising phase and redder in the falling one. Two $\gamma$-ray flares were detected by Fermi-GST during our monitoring: on the occasion of only one of them we found simultaneous optical brightening. A one-zone Synchrotron Self Compton (SSC) model appears too simple to explain the source behavior.
\\
\\
\noindent \textbf{Keywords}: Active Galactic Nuclei, Blazars, BL Lac objects.
\end{abstract}

\section{Introduction}
S5 1803+78 is a BL Lac object, a special class of Active Galactic Nuclei (AGN). BL Lacs take their names from the prototype source, discovered at Sonneberg in 1929 and then cataloged as the variable star BL Lacertae.
They are characterized by: a) a featureless optical continuum; b) large and fast variability; c) appreciable polarization; d) flat radio spectrum. Many BL Lacs are X-ray sources and are a substantial part of the extragalactic sources detected in $\gamma$ rays. 

The current model to explain the emission of BL Lacs is based on a supermassive black-hole, hosted in the galaxy center, which accretes matter from the surroundings. Part of the matter in the accretion disk is funnelled in a narrow relativistic jet, oriented at small angles with the observer's line of sight: the existence of such jets has been proven by VLBI radio observations. 

Most of the observed emission is believed to be produced in this jet by two different processes: 1) synchrotron radiation of relativistic electrons in a strong magnetic field; 2) inverse Compton radiation from these electrons on the ambient photons. Depending on the energy of the electrons and the intensity of the magnetic field, the peak of the synchrotron emission, in the $Log(\nu~F_{\nu}) ~vs ~Log(\nu)$ plane, may range between far IR and X rays. Accordingly, BL Lacs are classified as low frequency (LBL), intermediate frequency (IBL) and high frequency (HBL) peaked sources. The peak of the inverse Compton emission is between the hard X-ray and the $\gamma$-ray bands. During a flare the spectral energy distribution (SED) shape is deformed and generally shifted towards higher frequencies.

The light curves of several BL Lacs have been studied since their discovery, also using historical plate archives for long-term variability studies. Their behavior has proven to be generally quite erratic, save the case of OJ 287 [1], which has a periodicity of about 11 years for the major flares.  In a number of cases time scales of a few years have been found, though not strict periodicities (e.g. S5 0716+71 [2]; AO 0235+174 [3]; GB6 J1058+5628 [4]).
Long-term trends have also been found in a few sources, superposed on the short-term ones, extending for tens of years and possibly being just part of longer recurrency time scales (e.g. OQ 530 [5]; ON 231 [6]; S5 0716+71 [2]). The long-term variability may be due to intrinsic changes in the source or to geometrical effects: in this case the best explanation could be a precession of the relativistic jet, producing a variation of Doppler boosting and therefore an achromatic variation of the apparent source luminosity.

\section{S5 1803+78 in brief}

S5 1803+78 is a bright radio source discovered in 1981. Due to its circumpolar position it can be followed for a large part of the year. It has a large optical polarization and a redshift z=0.680, based on a single emission line assumed to be MgII (2900 \AA). It is well monitored in the radio band because it is a source of the ICRS reference frame and is also used as a geodetic reference source. It was observed by several X-ray satellites, but was not pointed by EGRET. Its first optical systematic monitoring was made by [7] in the years 1996-2001.
The overall SED is of the LBL type.

Most of the well-monitored LBL sources (e.g. BL Lacertae itself) show a correlation between optical luminosity and optical spectral slope, being Òbluer when brighterÓ.
On the contrary, S5 1803+78 showed very limited variation in the optical spectral slope, although it varied by more than 2 mag in flux.
We decided therefore to continue its monitoring to check if:
a) the color variation remained small in time;
b) the source showed any recurrent time scale in flux variations;
c) the source showed monotonic long-term trend.
The monitoring was performed with the telescopes of Asiago (183 cm), Loiano (152cm), Vallinfreda (50cm) and Greve in Chianti (30cm), mainly in the R band, but with several observations also in the B, V and I bands. The resulting light curve is shown in Fig. 1.

\section{Optical light curve}

The source underwent a number of flares during our monitoring: we derived by eye estimates of the starting and ending point for the rising and falling branches of each major well-sampled flare.
The luminosity variation rates range between 0.03 and 0.07 mag/day.
The average values of the rising and falling rates are not statistically different.
An inspection of the light curve (by eye and by FFT analysis) shows no strict periodicity for the flares. 
A possible time scale of 1300 days between major flares may be present.
\begin{figure*}
\centering
\includegraphics[scale=0.7]{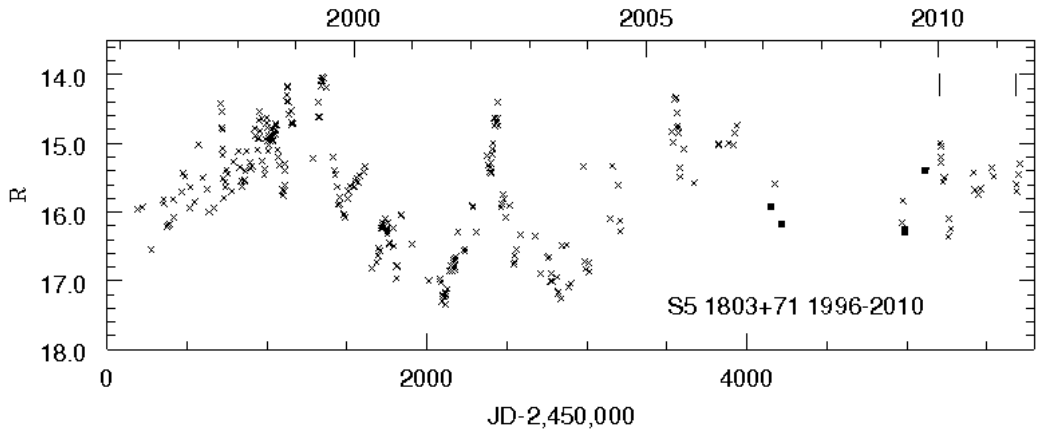}
\caption{Light curve in R band of S5 1803+78 from 1996 to 2010. Abscissa lower scale in Julian days, upper scale in years. The squares after JD 4000 are Swift-UVOT data extrapolated to the R band. Vertical dashes indicate the dates of Gamma-ray flares. }
              \label{LC}
\end{figure*}

The (V-I) color index showed small variations around the average value of V-I=1.1 mag during all our monitoring, a typical value for LBL sources, corresponding to spectral index -1.6.
The most accurate measures, carried out with the larger Asiago and Loiano telescopes during the strong flare at JD 3550, showed a bluer color during the rising phase and a redder one in the falling part.  Similar behavior was reported in the literature for a few other BL Lacs (S5 0716+71, 3C 66A [8]).
The UV spectral slope observed by Swift-UVOT was consistent with the optical one from our BVRI data, indicating a common origin (tail of the synchrotron emission) for optical and UV light.

\section{High energy observations}

The X-ray spectrum was observed by SWIFT on four occasions between 2007 and 2009: it always showed a flux level comparable to that observed by BeppoSAX in 1998, with small variations. 
Also the X-ray spectral slope (photon index) was stable, around +1.6, suggesting an Inverse Compton origin (see Fig.2).
During these X-ray observations the R magnitude was not very different, ranging between 15.4 to 16.3 mag: no clear correlation is evident between the X-ray and the optical  band.

   \begin{figure}
   \centering
   \includegraphics[width=7.0cm,angle=-90.0]{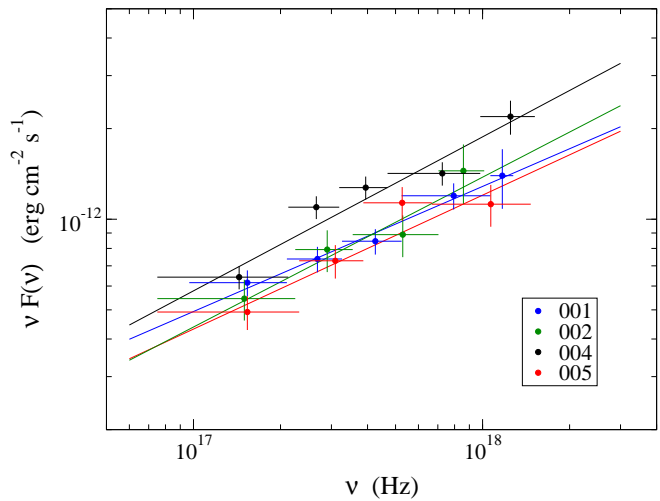}
      \caption{X-ray spectrum of S5 1803+78: different symbols are used for the four epochs. The numbers in the rectangle are the last digits of the observation ID in the {\it SWIFT} archive.}
         \label{XRT}
   \end{figure}

S5 1803+78 was first detected in $\gamma$ rays by Fermi-GST (1FGL [9]).  Its weekly sampled $\gamma$-ray light curve shows an average level around $2 \cdot 10^{-8} ph~cm^{-2} s^{-1}$ with oscillations within a factor of 2 ([10]). 
Our optical light curve has only a few points around the end of this 11-month time interval; the source was rather faint (R $\sim$ 16.0) and shows a short increase of 0.3 mag two weeks before a short $\gamma$-ray bump (MJD 54990). 

A strong flare, at $9 \cdot10^{-7} ph~cm^{-2} s^{-1}$ (E$\ge$100 MeV) on 2010 January 11 was reported by [11], 40 times brighter than the average value. 
We could observe the source in the optical 4 days after the $\gamma$-ray burst at R=14.9 mag with color index (V-I)=1.2 mag.  
It was already in the decreasing phase at a rate of 0.04 mag/day, a typical value for this source.
Extrapolating backwards, this trend to the epoch of the $\gamma$-ray flare gives R=14.7 mag for the peak value, rather lower than the maximum recorded flares of this source (R=14.0--14.2 mag).
A second brighter flare ($1 \cdot 10^{-6} ph~cm^{-2} s^{-1}$) was detected by Fermi-GST [12] on 2011 May 2: we observed the source in the following days, finding it at R=15.6 mag, at variance with the previous flare [13].  The source rose at about R=15.3 in the following 30 days. The dates of the flares are indicated in Fig. 1 with vertical dashes.

On the basis of these two contradictory Gamma-ray episodes it is rather premature to derive any strong conclusion on a correlation between optical and Gamma-ray flux density variations.

\section{Spectral energy distribution}

We used literature data, as well as our own observations and data analysis, to build the SED of the source, which is shown in Fig. 3. The Gamma-ray flux densities (violet dots) are from Fermi-GST: they have been retrieved from the ASI Science Data Center (ASDC) website and are the average values during the Aug-2008 Jul-2009 period.
The hard X-ray flux density at 20~keV (violet square) was calculated by converting the count rate in the 15--30 keV map obtained from the reduction of 54 months of BAT data using BATIMAGER software [14]. The conversion factor was derived from the Crab count rate, and its spectrum was used for calibration purposes, as explained in the BAT calibration status report\footnote{http://swift.gsfc.nasa.gov/docs/swift/analysis/bat\_digest.html}.  
The X-ray data in the 0.5 - 8 keV range  are those already reported in Fig.2, with the same color code.
Near UV data (blue dots) are from Swift-UVOT, simultaneous with one (blue) Swift-XRT pointing. The optical black dots are from SDSS. The optical and NIR red dots are from our observations: for clarity, only the maximum and minimum optical levels of the source during our monitoring are shown. Finally, mm and radio data are shown as black dots.

   \begin{figure}
   \centering
   \includegraphics[width=8cm,angle=-90.0]{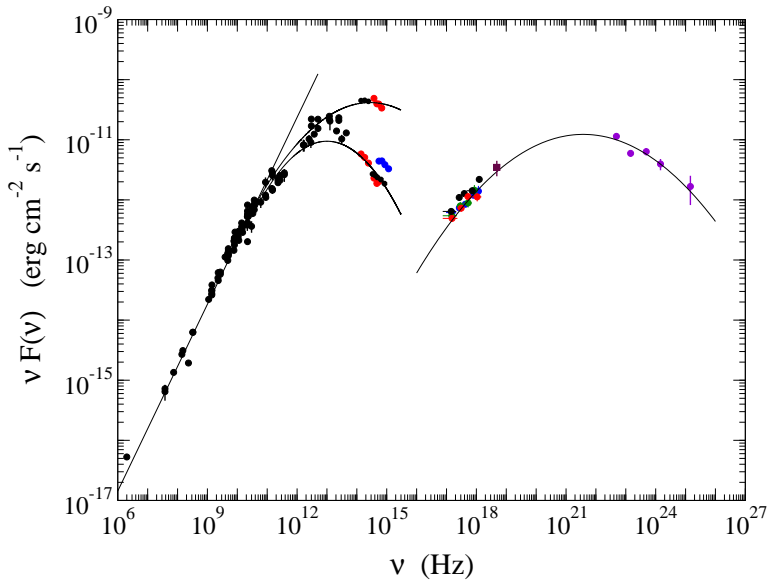}
      \caption{The SED of S5 1803+78 from non-simultaneous data. See the text for colour codes. The Log-parabolic fits to the Synchrotron component and Inverse Compton components are also shown.}
         \label{SED}
   \end{figure}
%

 The SED is that typical of LBL sources, with two large bell-shaped parts:
the synchrotron component, peaking between 10$^{13}$ and 10$^{14}$~Hz, is well fitted by a log-parabola [15]; 
the inverse Compton branch is well defined by the Swift-XRT, BAT and Fermi-GST instruments: a log-parabolic fit gives a peak at $3.8 \cdot 10^{21}$~ Hz, similar to that of other LBL objects observed by Fermi-GST. 
A bluer-when-brighter behaviour of the synchrotron component is quite evident.
The photon index in the $\gamma$-ray domain is 2.33, well within the range of values for LBL objects (2.21$\pm 0.16$, see [16]).

\section{Conclusions}

The optical light curve of S5 1803+78 shows several large (2 mag) flares with a possible time scale of about 1300 days. The flux variations are slow, with a typical rate of 0.04 mag/day, without a marked difference between the rising and falling branches. No monotonic long-term trend is apparent on a 14-year time span.
The optical spectral slope shows only small (10\%) variations even in large flares: accuracy of 0.01 mag is necessary to study these variations. On the occasion of a strong flare we had enough accuracy to detect  a bluer-when-rising behavior.
The overall SED is well described by two log-parabola, with peak positions and slopes typical of LBL sources.
The correlation between $\gamma$-ray and optical flares is based on two cases only: further observations are necessary to explore this topic, but a one-zone synchrotron self Compton model seems too simple to explain the behavior of this source.  If the flares are correlated, from the frequency of the optical flares we statistically expect at least one more $\gamma$-ray flare within the lifetime of Fermi-GST.

\end{document}